\documentclass[12pt]{article}
\usepackage{amsmath}
\usepackage{amstext,amssymb,amsfonts, graphicx,color,subfigure}

\advance\voffset by -1.5cm
\advance\hoffset by -1.25cm
\definecolor{darkgreen}{rgb}{0,0.6,0}
\def\thefootnote{\fnsymbol{footnote}}

\def\tq{\tilde q}

\def\be{\begin{equation}}
\def\ee{\end{equation}}
\def\ba{\begin{eqnarray}}
\def\ea{\end{eqnarray}}

\newcommand{\ddd}{\displaystyle}

\newcommand{\N}{{\cal N}}

\newcommand{\nn}{{\nonumber}}

\newcommand{\ttau}{{\tilde \tau}}

\def\<{\langle}
\def\>{\rangle}

\def\tmu{\tilde \mu}
\usepackage{color}

\usepackage[english, USenglish]{babel}
\usepackage{epsfig}
\usepackage{slashed}

\begin{document}


\thispagestyle{empty}
\renewcommand{\thefootnote}{\fnsymbol{footnote}}

{\hfill \parbox{3cm}{
 DESY 10-232 \\ 
}}

\bigskip\bigskip\bigskip

\begin{center} \noindent \Large \bf
Holographic dual of a boost-invariant 
plasma\\ with chemical potential
\end{center}

\bigskip\bigskip\bigskip

\centerline{ \normalsize \bf 
Tigran Kalaydzhyan\footnote[2]{\noindent \tt email: tigran.kalaydzhyan@desy.de}
and Ingo Kirsch\footnote[1]{\noindent \tt email: ingo.kirsch@desy.de}}

\bigskip\bigskip

\centerline{\it  DESY Hamburg, Theory Group,}
\centerline{\it Notkestrasse 85, 22607 Hamburg, Germany}
\vspace{0.3cm}

\bigskip\bigskip\bigskip

\bigskip\bigskip\bigskip

\renewcommand{\thefootnote}{\arabic{footnote}}

\centerline{\bf Abstract}
\medskip

{We construct a gravity dual of a boost-invariant flow of an $\N = 4$ $SU(N)$ supersymmetric
Yang-Mills gauge theory plasma with chemical potential. We present both a first-order corrected
late-time solution in Eddington-Finkelstein 
coordinates and a zeroth-order solution in parametric form in Fefferman-Graham coordinates. 
The resulting background takes the
form of a time-dependent AdS Reissner-Nordstr\"{o}m-type black hole whose horizons move into the bulk 
of the AdS space. The solution correctly reproduces
the energy and charge density as well as the viscosity of the plasma previously computed in the literature. }

\bigskip\bigskip

\newpage
\setcounter{tocdepth}{2}
\tableofcontents

\setcounter{equation}{0}
\section{Introduction}

In the recent years, the application of the AdS/CFT correspondence \cite{Maldacena} to 
the quark-gluon plasma (QGP) has become a very active research area.
One line of 
research within such holographic studies was
initiated by Janik and Peschanski \cite{Janik2005} who established a time-dependent
gravity dual of the boost-invariant flow of an $\N = 4$ plasma. 
This geometry has mainly been studied in the 
regime of large proper time, when the system is
near equilibrium and approaches the hydrodynamic regime (see however \cite{Janik09, Chesler}). In \cite{Nakamura}--\cite{Hellerthesis}
higher-order corrections to this late-time background were constructed and
found to be equivalent to a gradient expansion of hydrodynamics, see \cite{Janik:2010we}
for a review.

An important aspect of the plasma which has not yet received much attention in a time-dependent
gravity
background is the effects of chemical potentials, even though an asymptotic boost-invariant
geometry  ({\em without} corrections) 
dual to an $\N=4$ plasma with $U(1)$ $R$-charge is known for quite some time \cite{Bak}.
Also the transport coefficients of plasmas
with $U(1)$ currents have already been holographically computed
in \cite{Erdmenger}--\cite{Oz} (up to second order). 
Such currents are generated, for instance, 
shortly after the collision of two heavy ions, 
when the two sheets of color glass condensates have passed through each other and longitudinal color
electric and magnetic flux tubes are produced between the sheets \cite{McLerran}.
This gives rise to a large topological charge density $F_a^{\mu\nu} \tilde F^a_{\mu\nu}$, which in 
turn leads to an imbalance of the number of quarks
with left- and right-handed chirality and chemical potentials $\mu_R$ and $\mu_L$. 
In addition to the usual baryon chemical potential $\mu=\frac{\mu_R+\mu_L}{2}$, 
one may therefore also consider a chiral chemical potential $\mu_5=\frac{\mu_R-\mu_L}{2}$ 
which mimics the effect of an imbalanced chirality. 

In this paper we will construct a modification of the Janik-Peschanski background,
which will additionally include a time-dependent $U(1)$ gauge field.
The bulk theory will be five-dimensional Einstein-Maxwell gravity with a negative cosmological constant and a Chern-Simons term. As in the case without chemical potential, it appears to be difficult
to find an analytic solution for all times and we will restrict to solving the equations
of motion at late times. 
As a further simplification, we seek for a solution in which 
only the time-component of the $U(1)$ gauge field dual to the chemical potential 
is non-vanishing (the spatial components are set to zero). 
Asymptotically, at large proper time $\tau$, we may expand
the late-time geometry in powers of $\tau^{-2/3}$. 
Employing both 
Eddington-Finkelstein and Fefferman-Graham coordinates we present
the late-time solution up to first order (in $\tau^{-2/3}$). The
resulting background will essentially take the form of a time-dependent $AdS_5$ 
Reissner-Nordstr\"{o}m solution whose inner and outer horizon move into the bulk of the
AdS space. This background can be extended to a full type IIB supergravity solution (by taking the product with an $S^5$) and is dual to a strongly-coupled $\mathcal{N} = 4$ $SU(N)$ supersymmetric-Yang-Mills plasma with a non-vanishing chemical potential.

\setcounter{equation}{0}
\section{Late-time background in Eddington-Finkelstein coordinates}

In this section we are interested in finding a late-time gravity dual of an
expanding $\mathcal{N} = 4$ viscous plasma with non-vanishing chemical potential.

The relevant five-dimensional Einstein-Maxwell-Chern-Simons action is given by
\begin{align}\label{action}
S=\frac{1}{16\pi G_5}\!\int\!d^5x\, \sqrt{-g_5} \left(R + 12 -
F_{\alpha\beta}F^{\alpha\beta} + \frac{4\kappa}{3}\epsilon^{\sigma\alpha\beta\gamma\delta}A_{\sigma} F_{\alpha\beta} F_{\gamma\delta}
\right),
\end{align}
where $\alpha, \beta,...$ denote the 5D bulk coordinates.
The cosmological constant is $\Lambda=-6$ and the Chern-Simons parameter is fixed 
as $\kappa= -1/(2\sqrt{3})$. Also, $1/(16\pi G_5)=N_c^2/(8\pi^2)$ for an $\N=4$ plasma \cite{Erdmenger}.
 The corresponding equations
of motion are given by the combined system of Einstein-Maxwell equations,
\begin{align} \label{EM}
R_{\alpha\beta}-\frac{1}{2}g_{\alpha\beta} R - 6  g_{\alpha\beta} + 
2\left(F_{\alpha}{}^{\gamma} F_{\gamma\beta} -
\frac{1}{4} g_{\alpha\beta} F^2\right) = 0 \,,
\end{align}
and covariant Maxwell equations (with Chern-Simons-term),
\begin{align}\label{EM2}
 \nabla_\beta F^{\beta\alpha} + \kappa\epsilon^{\alpha\beta\gamma\delta\sigma} F_{\beta\gamma} F_{\delta\sigma} = 0 \,.
\end{align}
$F^{\alpha\beta}$ is the field strength of the $U(1)$ gauge field $A_\alpha$
we wish to introduce in the background.

\subsection{Boosted black brane solution}

Our starting point for the construction of a time-dependent solution is the {\em static} $AdS_5$ Reissner-Nordstr\"om (RN) black-hole solution \cite{Chamblin}. Using ingoing Eddington-Finkelstein
coordinates, we may write the RN metric and gauge field as
\begin{align}
ds^2 &= - r^2 (1-\frac{m}{r^4} + \frac{q^2}{r^6}) dv^2 +
2dv dr+ r^2 d\vec x^2\,,\label{static}\\
A&=-\frac{\sqrt{3} q}{2r^2} d\ttau \,,
\end{align}
with mass $m$ and charge $q$. Here $v$ is a time-like coordinate (not to be mixed up with 
the scaling variable $v$ introduced below), $\vec x$ are the spatial coordinates
on the boundary, and $r$ parameterizes the holographic direction.
The location of the outer horizon 
$r_+=r_+(m,q)$ is given by the largest real positive 
root of $V(r_+)=r_+^6-mr_+^2+q^2=0$. 

A charged black hole is dual to a fluid at finite
temperature $T$ and chemical potential~$\mu$.
Both the Hawking temperature and the chemical potential are given in
terms of $r_+$ by~\cite{Chamblin}
\begin{align}
T=-\frac{1}{4\pi}
g_{vv}'(r_+)\,,\qquad \mu = \frac{\sqrt{3}q}{2 r_+^2}\,.
\end{align}
These relations can be inverted to give $m$ and $q$ as functions of $T$ and $\mu$
\cite{Erdmenger},
\begin{align}
m&=r_+^4  \frac{3\gamma-1}{\gamma+1}\,,
\qquad
q=\frac{2\mu}{\sqrt{3}} r_+^2 \label{mq} \,,
\end{align}
with 
\begin{align}\label{rplus}
r_+ = \frac{\pi T}{2} (\gamma+1)\,,\qquad \gamma=\sqrt{1+\frac{8\mu^2}{3\pi^2 T^2}}\,.
\end{align} 

\medskip
Following \cite{Hubeny, Heller}, we now consider the corresponding five-dimensional boosted
charged black brane solution given by
\begin{align}
ds^2&=-2 u_\mu dx^\mu dr - r^2 \left(1-\frac{m}{r^4} + \frac{q^2}{r^6}\right)
u_\mu u_\nu dx^\mu dx^\nu+r^2 P_{\mu\nu} dx^\mu dx^\nu \,,\nn\\
A&= \frac{\sqrt{3} q}{2 r^2} u_\mu dx^\mu\,, \qquad P_{\mu\nu}=\eta_{\mu\nu}+u_\mu u_\nu \,,\label{boostedsol}
\end{align}
where $u^\mu$ is the boost velocity along $x^\mu$ ($\mu=0,1,2,3$), and $m=m(\mu,T)$ and 
$q=q(\mu,T)$ as given by (\ref{mq}). From this solution we may deduce
a time-dependent solution by choosing the frame $u^\mu=(1,0,0,0)$ and 
introducing an Eddington-Finkelstein proper time-like coordinate $\tilde \tau$
and rapidity-like coordinate $y$. We also substitute the
asymptotic late-time behaviour of $T$ \cite{Bjorken} and $\mu$, 
\begin{align}
\label{T_and_mu_scaling}
 T = \Lambda \ttau^{-1/3} \qquad \textmd{and} \qquad \mu=\tmu_0 \ttau^{-1/3}\,,\qquad \Lambda,\tmu_0=const.\,,
\end{align}
into the explicit expressions for $m$ and $q$. Here we assumed $\mu \propto T$, as
one would expect for a perfect fluid, such that the quotient $\mu/T=\tmu_0/\Lambda=const.$ 
is independent of time. This leads to the
following metric\footnote{There is an additional $1$ in the factor $(1+r \ttau)^2$
in front of $dy^2$ which is not expected from the boosted solution
(\ref{boostinv}). This is to ensure an asymptotic AdS space in the limit
$\Lambda \rightarrow 0$, see \cite{Kinoshita} for details.}
\begin{align}
ds^2&=- r^2 (1-\frac{m(\ttau)}{r^4} + \frac{q(\ttau)^2}{r^6}) d\ttau^2 + 
2d\ttau dr + (1+r \ttau)^2 dy^2 + r^2 dx_\perp^2\,, \nn\\
A&= -\frac{\sqrt{3} q}{2 r^2} d\ttau \,, 
\label{boostinv}
\end{align}
with coefficients 
\begin{align}
m(\ttau)&=b(\ttau)^{-4}\equiv r_+(\ttau)^4 \frac{3\gamma-1}{\gamma+1}\,,\quad
q(\ttau)=\frac{2\tmu_0}{\sqrt{3}\ttau^{1/3}} r_+(\ttau)^2 \label{m_definition}\,,\\
r_+(\ttau) &= \frac{\pi \Lambda}{2\ttau^{1/3}} (\gamma+1) \,,\quad \gamma=\sqrt{1+\frac{8\tmu_0^2}{3\pi^2 \Lambda^2}}\,. \label{horizon}
\end{align}
For $q = 0$ (or $\tmu_0 = 0$), this metric reduces to the uncharged (zeroth-order) late-time solution
in Eddington-Finkelstein coordinates found in \cite{Heller, Kinoshita, Hellerthesis} ($m = b^{-4} = \pi^4 \Lambda^4 \ttau^{-4/3}$ there). Note that the size of the outer (and inner) horizon $r_+$
($r_-$) decreases with time.

\subsection{Zeroth-order solution and first-order correction}
The boosted metric (\ref{boostinv}) is not an exact solution of the Einstein-Maxwell
equations. It is a good approximation of the boost-invariant solution
at large $\ttau$ though. At smaller $\ttau$, it receives subleading 
corrections corresponding to higher-order gradient corrections to the energy-momentum tensor
and $U(1)$ current, which will be discussed in section~\ref{secTransport}. These corrections to the metric (\ref{boostinv}) can be
found by choosing the following metric ansatz for the time-dependent
solution:\footnote{For this particular ansatz, 
the Maxwell equation reduces to 
$\frac{1}{\sqrt{-g}} \partial_\beta (\sqrt{-g} F^{\beta\alpha})=0$. The Chern-Simons
term is absent, since only $F_{r\ttau}$ and $F_{\ttau r}$ are non-vanishing.}
\begin{align}
\label{ansatz}
ds^2&=- r^2 e^{a(\ttau,r)} d\ttau^2 + 2d\ttau dr + 
(1+r \ttau)^2 e^{b(\ttau,r)} dy^2 + r^2 e^{c(\ttau,r)}  dx_\perp^2 \,, \nonumber \\ 
A&= d(\ttau,r) d\ttau \,.
\end{align}
As in the case without chemical potential, we may introduce
the scaling variable $v=r\ttau^{1/3}$ and expand the metric coefficients
in powers of $\ttau^{-2/3}$,
\begin{align}
e^{a(\ttau,r)} &=A(v) + a_1(v) \ttau^{-2/3} + \ldots\,,\\
e^{b(\ttau,r)} &=B(v) \exp({b_1(v) \ttau^{-2/3} + \ldots})\,,\\
e^{c(\ttau,r)} &=C(v) \exp({c_1(v) \ttau^{-2/3} + \ldots})\,. 
\end{align}
Similarly, for the coefficient of the gauge field we choose
\begin{align}
d(\ttau,r) = D(v) \ttau^{-1/3} \exp({d_1(v) \ttau^{-2/3} + \ldots}) \,. \label{endofansatz}
\end{align}
Note that the gauge field has an overall factor $\ttau^{-1/3}$. The existence
of a late-time scaling variable $v$ will be shown in section~\ref{seclatetime}.

\medskip
The system of Einstein-Maxwell equations (\ref{EM}) and Maxwell equations (\ref{EM2})  
can then be solved order by order in
$\ttau^{-2/3}$. At zeroth-order in $\ttau^{-2/3}$, we find
the coefficients 
\begin{align}
A(v)=1-\frac{m_0}{v^4} + \frac{\tq_0^2}{v^6} \,,\qquad
B(v)=C(v)=1\,,\qquad
D(v)=-\frac{\sqrt{3} \tq_0}{2 v^2 } \,, \label{zerocoeff}
\end{align}
where we defined the time-independent variables
\begin{align}
m_0=b_0^{-4}\equiv\ttau^{4/3} m(\ttau) \,,\qquad \tq_0=\ttau q(\ttau)
\label{mbq}
\end{align}
with $m(\ttau)$ and $q(\ttau)$ as in (\ref{m_definition}). In the same way, we 
also define the variable
\begin{align}
\zeta_+ = r_+(\ttau)\ttau^{1/3} 
\end{align}
from the (outer) horizon $r_+$ as given by (\ref{horizon}). $A(v),...,D(v)$ are 
in agreement with the metric (\ref{boostinv}) deduced from the boosted black brane.

\medskip
At first order in $\ttau^{-2/3}$, we find the coefficients
\begin{equation}
\label{firstcoeff}
\begin{array}{ l }
\ddd a_1(v) = -\frac{4 q^2_0}{3 v^7}+\frac{2 m_0}{3 v^5}+\frac{C_2}{v^4} \,, \\\\
\ddd b_1(v) = -2c_1(v) = -\frac{4}{3 v}+C_3+\frac{1}{6} \sum\limits_{i=1}^6\frac{3 C_1 \log[v-\zeta_i]-4\log[v-\zeta_i] \zeta_i^3}{3 \zeta_i^4 - m_0} \,,\\\\
\ddd d_1(v) = -\frac{2}{3 v}+\frac{1}{2} v^2 C_1 \,,
\end{array}
\end{equation}
where $\zeta_i$ are the solutions of
\begin{align}
\label{roots}
\zeta_i^6 -m_0 \zeta_i^2 + \tq_0^2 = 0 \,.
\end{align}
The resulting expression for $b_1(v)$ is real, even though
we need to consider all six roots of (\ref{roots}) including the imaginary ones. Explicit expressions for these roots can be found in appendix~\ref{appA}.  Note that one of the six roots of this equation corresponds to the outer horizon $\zeta_+$. In Reissner-Nordstr\"om solutions there is
always an upper bound on the charge~$\tq_0$, at which the discriminant of the 
equation (\ref{roots}) vanishes,
\begin{align}
\tq_0 \leq \tq_0^{extr.} = \sqrt[4]{\frac{4}{27}m_0^3}\,.  \label{boundEF}
\end{align}
For larger values of $\tq_0$, there would be a naked singularity at the origin. Remarkably, this bound
is satisfied for any value of the quotient $\tmu_0/\Lambda$ and saturated in the
limit $\tmu_0/\Lambda \rightarrow \infty$, as can be seen by substituting (\ref{mbq}) with (\ref{m_definition}) into the bound (\ref{boundEF}). In other words, there is {\em no} 
bound on the chemical potential. Nevertheless, let us assume that
$\tmu_0 \ll \Lambda$ in order to avoid potential stability problems \cite{Gubser}, which 
arise when the black hole is close to extremality.

We still need to fix the integration constants $C_{1,2,3}$. $C_1$ can be found by requiring
regularity of the first-order solution (\ref{firstcoeff}) at the outer horizon,
{\em i.e.}\ $C_1$ should be a function of the positive root $\zeta_+$. More precisely,
by choosing
\begin{align}
C_1 = \frac{4}{3}\zeta_+^3 \,,
\end{align}
we cancel the $\log[v-\zeta_+]$ terms in $b_1(v)$, which are singular at $v=\zeta_+$. 
The metric then still contains singularities but
they are hidden behind the outer horizon. 

The constant $C_3$ is fixed by the requirement that the metric reduces to
a pure AdS space in the limit $\Lambda \rightarrow 0$. This simply sets $C_3$ to zero,
\begin{align}
 C_3=0 \,.
\end{align}
There is one remaining integration constant $C_2$ which can not be fixed at first order. 
Note that, in general, at each order $k$ there is one integration constant which can only be fixed 
by regularity at order $k+1$ \cite{Heller}, $C_2$ in our case. Nevertheless, we may guess the 
correct value for $C_2$ by comparing with the uncharged solution \cite{Heller, Kinoshita, Hellerthesis}, 
in which $C_2=\frac{2}{3} \zeta_H^3= \frac{2}{3} \pi^3 \Lambda^3$. As for $C_1$, it seems natural to replace the horizon $\zeta_H$ of the uncharged solution by the outer horizon $\zeta_+$ of the charged solution such that 
\begin{align}
C_2=\frac{2}{3} \zeta_+^3 \,.
\end{align}
Later in section~\ref{secTransport} we will justify this value again. It will turn out to 
correctly reproduce the expected transport coefficients. 

We have checked that for $\tq_0=0$ (or, equivalently, $\tmu_0=0$) 
the metric reduces to the first-order corrected uncharged solution found in
\cite{Heller, Kinoshita, Hellerthesis}. Moreover, for the Kretschmann scalar we find
\begin{align}
R^2_{\mu\nu\rho\sigma} 
&= \frac{4(127\tq_0^4-90m_0 \tq_0^2 v^2 + 18m_0^2v^4+2\tq_0^2v^6+10v^{12})}{v^{12}} \\
&~~~+ \frac{8(254 \tq_0^4-150m_0\tq_0^2v^2+24m_0^2v^4+2\tq_0^2v^6-45\tq_0^2v^3 C_2 +18m_0 v^5 C_2)}{v^{13}} \ttau^{-2/3}+ ... \,,\nonumber
\end{align}
which is only singular at $v=0$. In the limit $\tq_0 \rightarrow 0$, 
we have $m_0\rightarrow \pi^4 \Lambda^4$ and $R^2_{\mu\nu\rho\sigma}$ reduces to the
corresponding expression in the uncharged case, see \cite{Kinoshita}.

We have thus constructed a natural extension of the first-order corrected boost-invariant plasma
geometry of \cite{Heller, Kinoshita,Hellerthesis} to the corresponding one with non-trivial $U(1)$ gauge field.

\subsection{Transport coefficients from the background}\label{secTransport}

In the hydrodynamic approximation, the energy-momentum and $U(1)$ current are given by
\begin{align} \label{TJ}
\langle T_{\mu\nu} \rangle = \frac{\varepsilon}{3} (4 u_\mu u_\nu
+\eta_{\mu\nu}) + \Pi_{\mu\nu} \,,\qquad
\langle J_\mu \rangle = \rho u_\mu + \Upsilon_\mu \,,
\end{align} 
where the first terms on the right hand side correspond to a perfect fluid with
chemical potential. Since the velocity field $u^{\mu}$, energy density $\varepsilon$
and charge density $\rho$ vary slowly with the spacetime coordinates, the energy-momentum
tensor and current receive higher-order gradient corrections given by (up to first 
order)
\begin{align} \label{correc}
\Pi_{\mu\nu} = - \eta \sigma_{\mu\nu} \,,\qquad
\Upsilon_\mu=-\sigma P_\mu{}^\alpha \partial_\alpha \frac{\mu}{T}
+\xi \epsilon_\mu{}^{\rho\sigma\tau} u_\rho\partial_\sigma u_\tau\,,
\end{align}
where $\eta$, $\sigma$ and $\xi$ denote the viscosity, conductivity and vorticity coefficient,
respectively. The corrections satisfy $u^\nu \Upsilon_\nu=0$ and $u^\nu \Pi_{\mu\nu}=0$. 
The transport coefficients of the fluid entering
these corrections were holographically computed in \cite{Erdmenger, Banerjee, Yee}
(up to second order) by slowly varying $u^{\mu}$, $q$ and $m$ in the boosted solution 
(\ref{boostedsol}) with the space-time coordinates $x^\mu$. In this way the 
hydrodynamic equations are obtained from AdS/CFT without constructing
an explicit  solution. 

In the following we will compute the first-order corrections directly from our time-dependent solution
using holographic renormalization techniques \cite{Bianchi:2001kw}. Recently, a rigorous holographic renormalization of
the Einstein-Maxwell-Chern-Simons theory, including the full back-reaction of the gauge field, 
has been performed in \cite{Yee2}. The energy-momentum tensor can be obtained 
from
\begin{align}
\label{T_definition}
\langle T_{\mu\nu} \rangle=\lim\limits_{r\rightarrow\infty} \left[ \frac{N_c^2}{4\pi^2}r^{2}\left(K_{\mu\nu}-K \gamma_{\mu\nu}-3\gamma_{\mu\nu}+\frac{1}{2}G_{\mu\nu}\right) \right] \,,
\end{align}
where $\gamma_{\mu\nu}$ is the induce metric on a constant-$r$ hypersurface, which regularizes the boundary. $K_{\mu\nu}$ is the extrinsic curvature of on this hypersurface, $K$ the corresponding scalar $K=K_{\mu\nu}\gamma^{\mu\nu}$ and $G_{\mu\nu}$ the boundary Einstein tensor with respect to the metric $\gamma_{\mu\nu}$. Substituting our explicit first-order solution into (\ref{T_definition}), we find the time-dependent energy density\footnote{Asymptotically, $\ttau$ can be identified with the
proper time $\tau$, $\ttau \approx \tau$, see section~\ref{sec34} below.}
\begin{align}
\varepsilon(\ttau) = \frac{\varepsilon_0}{\ttau^{4/3}} -  \frac{2\eta_0}{\ttau^{2}} \,,
\end{align}
with
\begin{align}
 \quad  
 \varepsilon_0\equiv\frac{3N_c^2}{8 \pi^2 b^4_0} \,,\qquad
  \eta_0 \equiv  \frac{3N_c^2}{16 \pi^2} C_2 =\frac{N_c^2}{8 \pi^2}
  \zeta_+ ^3 \label{eps}\,,
\end{align}
and $b_0=b\ttau^{-1/3}$ as in (\ref{mbq}), see appendix \ref{extrinsic_curvature} for more details on the computation. The first term in $\varepsilon(\ttau)$ is the zeroth-order energy density and is in agreement with 
that in \cite{Erdmenger}, see Eq.~(20a) therein. The second term is the
first-order correction and formally agrees with that in the uncharged case \cite{Heller,Kinoshita,Hellerthesis}
but now with a more general shear viscosity
$\eta_0=\eta_0(\tmu_0,\Lambda)$.\footnote{In order to check $\eta_0$ in the limit of vanishing
chemical potential, we note that the viscosity is differently
normalised in \cite{Kinoshita}. Consider
$\frac{\eta_0}{\varepsilon_0}=\frac{1}{3}{\zeta_+^3 b_0^4}\rightarrow
\frac{1}{3\pi\Lambda}$ for $\tmu_0=0$. This is identical with $\eta_0^{KMNO}=\frac{1}{3w}$ found in
\cite{Kinoshita} since $\pi \Lambda=w$ there.} This correction is also
in exact agreement with the first-order gradient correction to the 
energy-momentum tensor computed in \cite{Erdmenger}. There \cite{Erdmenger}, the viscosity
was found to be
\begin{align}
\eta= \frac{s}{4\pi}
= \frac{N_c^2}{8\pi^2} r_+^3=\frac{N_c^2}{8\pi^2} \zeta_+^3 \ttau^{-1} \,,
\end{align}
with $r_+$ as in (\ref{rplus}) (The $\N=4$ plasma saturates the KSS bound \cite{KSS}). Here we have already substituted the asymptotic behaviour
$T=\Lambda \ttau^{-1/3}$ and $r_+=\zeta_+\ttau^{-1/3}$. Given that $\eta_0$ is defined as $\eta_0=\eta \ttau$, we get the same $\eta_0$ as in (\ref{eps}) and thus agreement with \cite{Erdmenger}. 

Similarly, the expectation value of the R-charge current can be computed from
\begin{align}
\langle J^{\mu} \rangle =  \frac{N_c^2}{4 \pi^2}\left( \eta^{\mu\rho}A^{(2)}_{\rho}
- \frac{\kappa}{2} \epsilon^{\mu\nu\sigma\rho} A^{(0)}_\nu F^{(0)}_{\sigma\rho}
\right) \,,\label{J_definition}
\end{align}
where $A^{(n)}_\rho$ is the $r^{-n}$ coefficient of the large-$r$ expansion of 
the gauge field $A_\rho$. Since the spatial components of the gauge field are zero, 
the second term proportional to $\kappa$ is absent in our case. 
Substituting the solution for the gauge field into (\ref{J_definition}), we
read off the $U(1)$ charge density
\begin{align} \label{chdensityasymp}
 \rho(\ttau) = \frac{N_c^2}{4 \pi^2} \frac{\sqrt{3}\tq_0}{2} \frac{1}{\ttau} 
\end{align}
with $\tq_0=q \ttau$ as in (\ref{mbq}). Recalling $\sqrt{3} q/2=\mu r_+^2$, we find agreement
with the zeroth-order charge density in \cite{Erdmenger}, see Eq.~(20b) therein.
The asymptotic $1/\ttau$ behaviour of the charge density was also found in \cite{Bak}.
There are no first-order corrections to the charge density in our case.

More generally, for gauge fields with vanishing spatial components, there are no 
higher-order gradient corrections. This follows directly from the
relation $u^\nu \Upsilon_\nu=0$. The corrections $\Upsilon_\nu$ are 
orthogonal to $u^\nu$ and cannot come from the near boundary expansion of 
a gauge field proportional to $u^\nu$.

\setcounter{equation}{0}
\section{Late-time solution in Fefferman-Graham coordinates}

In this section we seek for a time-dependent solution of the Einstein-Maxwell equations 
(\ref{EM}) and (\ref{EM2}) in Fefferman-Graham coordinates. 

\subsection{General ansatz and near-boundary behaviour}

In Fefferman-Graham coordinates, we choose the same metric ansatz as in the uncharged
case \cite{Janik2005} given by
\begin{align}
ds^2 = \frac{1}{z^2} \left( -e^{a(\tau, z)}d\tau^2 + e^{b(\tau, z)} \tau^2 dy^2
+e^{c(\tau, z)} dx_{\perp}^2 + dz^2 \right) \label{ansatzmetric} \,.
\end{align}
Of course, the warp factors $a(\tau, z)$, $b(\tau, z)$ and $c(\tau, z)$ will be modified
due to the effects
from the back-reaction of the gauge field. As before, we set the spatial components of the
gauge field to zero and assume a non-vanishing time-component,
\begin{align}
\label{ansatzA}
A_0=-d(\tau,z) \,,\qquad A_y=A_z=A_{x_\perp}=0 \,.
\end{align}

Let us first study the general behaviour of the solution near the boundary at $z = 0$.
Following \cite{Janik09}, we choose the small-$z$ expansions
\begin{align}    
a(\tau, z) &= -\varepsilon(\tau) z^4 + a_{6}(\tau) z^{6}+ a_{8}(\tau)z^{8}+...\,, \nn\\ 
b(\tau, z) &= b_4(\tau) z^4 + b_{6}(\tau)z^{6}+ b_{8}(\tau)z^{8}+... \,, \nn\\ 
c(\tau, z) &= c_4(\tau) z^4 + c_{6}(\tau)z^{6}+ c_{8}(\tau)z^{8}+... 
\end{align}
and
\begin{align}
d(\tau,z)= \rho(\tau) z^2 + d_4(\tau) z^4+ d_6(\tau) z^6 + ...
\end{align}
Here the lowest coefficients are determined by the energy and charge density, respectively.
For instance, solving the Einstein-Maxwell equations to lowest order in $z$, we obtain
\begin{align} 
b_4(\tau)=-\left(\varepsilon(\tau) + \tau \varepsilon'(\tau)\right), \quad c_4(\tau)=\varepsilon(\tau) + \textstyle\frac{1}{2} \tau \varepsilon'(\tau)\,,
\end{align}
as in \cite{Janik09}. There is no back-reaction of the gauge field on the geometry at this order
($\rho(\tau)$ does neither appear in $b_4(\tau)$ nor $c_4(\tau)$).
Likewise, the metric does not enter the Maxwell equations at this order.  However, other than
the energy density $\varepsilon(\tau)$, which can be freely chosen (at least at early times), 
the charge density $\rho(\tau)$ is uniquely fixed by the $z$-component of the Maxwell equations,
\begin{align}
 -\ddd\frac{2\rho(\tau)}{\tau}-2\rho'(\tau)=0 \,,
\end{align}
which is solved by
\begin{align}
\rho(\tau) = \frac{q_0}{\tau} \label{chdensity}\,,
\end{align}
$q_0=const$. Any dependence on the warp factors has dropped out in the Maxwell equations such
that $\rho(\tau)$ is independent of $\varepsilon(\tau)$. The result (\ref{chdensity}) for the
charge density holds for all times $\tau>0$. Remarkably, the charge density diverges at 
$\tau=0$.\footnote{Generic solutions of viscous fluid dynamics are not expected to be regular
in the infinite past (see footnote~4 in \cite{Erdmenger} in this context): The 
volume element on the boundary at constant proper time scales linearly with $\tau$.
Integrating the charge density ($\propto 1/\tau$) over this volume element ($\propto \tau$) 
yields a constant total charge. Thus, even though the charge density is divergent, 
the total charge is regular, even at $\tau=0$, ensuring the validity of the hydrodynamic
approximation.\label{footnote5}}
 
Solving the system of equations (\ref{EM}) and (\ref{EM2}) order by order, we find the solution
up to order $z^8$,
\begin{align}\label{warpfactors}
a(\tau, z) &= -\varepsilon(\tau) z^4 + \left(-\frac{\varepsilon'(\tau)}{4\tau}-\frac{\varepsilon''(\tau)}{12}+\frac{10 \rho(\tau)^2}{9} \right)z^6\nn\\
&~~~-\left( \frac{1}{6} \varepsilon (\tau )^2+\frac{1}{6} \tau  \varepsilon'(\tau ) \varepsilon (\tau ) + \frac{1}{16} \tau ^2 \varepsilon'(\tau)^2-\frac{\varepsilon'(\tau )}{128 \tau ^3}+\frac{\varepsilon''(\tau)}{128 \tau ^2}\right.\nn\\
&~~~~~~\left.+\frac{\varepsilon^{(3)}(\tau )}{64 \tau }+\frac{1}{384}\varepsilon^{(4)}(\tau ) + \frac{\rho(\tau)^2}{36\tau^2} \right) z^{8} + ... \nn\,,\\
d(\tau, z) &= \rho(\tau) \left(z^2 - \frac{\varepsilon(\tau)}{3} z^6 + \left( \frac{2\rho(\tau)^2}{9} - \frac{\varepsilon'(\tau)}{16\tau} - \frac{\varepsilon''(\tau)}{48} \right)
z^8 + ...\right)
\end{align}
and similar expressions for $b(\tau,z)$ and $c(\tau,z)$. 
These expressions for the warp factors generalise the corresponding ones for $q_0=0$
found in \cite{Janik09}. They describe the all-time near boundary
behaviour of the background as a function of the energy and charge density.

\subsection{Late-time ansatz for the background} \label{seclatetime}

A full analytical all-time solution is difficult to find, even in the uncharged case ($q_0 = 0$). 
It is however possible
to find a late-time solution. The general late-time behaviour of the energy and charge
densities can be found as follows (For the energy density the derivation is very similar
to that in \cite{Janik2005, Nakamura}).
In the local rest frame the energy-momentum tensor is diagonal with elements $T_{\tau\tau}$,
$T_{yy}$ and $T_{xx}=T_{x_2x_2}=T_{x_3 x_3}$ and the current has only a
time-component $J_\tau$ while $J_y=J_{x_2}=J_{x_3}=0$. Moreover, we assume that these components
depend only on~$\tau$. 

Using proper time and rapidity coordinates in flat Minkowski spacetime, 
defined by $x_0=\tau \cosh y$ and $x_1=\tau \sinh y$, 
\begin{align}
ds^2 = -d\tau^2 +  \tau^2 dy^2+ dx_{\perp}^2, 
\end{align}
the tracelessness condition $T^\nu{}_\nu=0$, energy-momentum conservation $T^{\mu\nu}{}_{;\nu}=0$ 
and charge conservation $J^\nu{}_{;\nu}=0$ have the form
\begin{align}
-  T_{\tau\tau} + \frac{1}{\tau^2} T_{yy} +T_{xx} = 0\,, \\
\tau \partial_\tau T_{\tau\tau} + T_{\tau\tau} + \frac{1}{\tau^2} T_{yy} = 0\,, \\
\tau \partial_\tau J_\tau + J_\tau = 0 \,.
\end{align}
Here we assumed that the anomaly in the $U(1)$ current is absent, which is true
for our simple ansatz of the gauge field.

Comparing with the zeroth-order energy-momentum tensor and current
given in (\ref{TJ}), in the frame $u^\nu=(1,0,0,0)$ we obtain
\begin{align}
\varepsilon(\tau) = \frac{\varepsilon_0}{\tau^{4/3}} \,,\qquad
\rho(\tau) = \frac{q_0}{\tau}\,. \label{asymp}
\end{align}
We observe that the asymptotic charge density (\ref{asymp}) is in exact agreement 
with the expression (\ref{chdensity}) for the charge density, which is valid for all times. In other
words, the late time charge density (\ref{asymp}) does not receive any higher-order gradient
corrections, in agreement with our findings in the previous section.

Substituting the asymptotic behaviour (\ref{asymp}) into the general solution (\ref{warpfactors})
  and expanding the resulting expressions for large $\tau$, we get ($\varepsilon_0=1$)
\begin{align}
a(\tau, z) &=-\frac{z^4}{\tau ^{4/3}}+\frac{2+30 {q_0}^2 \tau ^{4/3}}{27 \tau ^{10/3}}z^6+\frac{10-27 {q_0}^2 \tau ^{4/3}-54 \tau ^{8/3}}{972 \tau ^{16/3}}z^8 + \cdots\,,\\
b(\tau, z) &=\frac{z^4}{3 \tau ^{4/3}}-\frac{14+18 \tau ^{4/3} q_0^2}{81 \tau ^{10/3}}z^6 +\frac{-130+243 \tau ^{4/3} q_0^2-162 \tau ^{8/3}}{2916\tau ^{16/3}}z^8+ \cdots\,,\\
c(\tau, z) &=\frac{z^4}{3 \tau ^{4/3}}+ \left(\frac{4}{81 \tau ^{10/3}}-\frac{2 q_0^2}{9 \tau ^2}\right)z^6+\frac{ 50-81 \tau ^{4/3} q_0^2-162 \tau ^{8/3}}{2916 \tau ^{16/3}}z^8+ \cdots\,,
\end{align}
\begin{align}
\label{dtauz}d(\tau, z) &=\frac{q_0}{\tau }z^2-\frac{q_0}{3 \tau ^{7/3}}z^6+ \left(\frac{q_0}{54 \tau ^{13/3}}+\frac{2{q_0}^3}{9 \tau ^3}\right)z^8 + \cdots\,\,.
\end{align} 
We find that the dominant terms at large $\tau$ scale as
\begin{align}
a_n(\tau) z^n \sim \frac{z^n}{\tau^{n/3}} \,,\qquad 
d_n (\tau) z^n \sim  \frac{1}{\tau^{1/3}} \frac{z^n}{\tau^{n/3}} \,,
\end{align}
and similarly $b_n(\tau) z^n$ and $c_n(\tau) z^n$. As in \cite{Janik2005}, it is therefore
useful to introduce the scaling variable\footnote{With hindsight, this justifies the 
introduction of the scaling variable $v=r \ttau^{1/3}$ in the previous section
for the late-time solution in 
Eddington-Finkelstein coordinates.} 
\begin{align}
v=\frac{z}{\tau^{1/3}} \, .
\end{align}

This suggests the following ansatz at late times,
\begin{align}
a(\tau, z) &= a_0(v) +  a_1(v) \frac{1}{\tau^{2/3}} + ...\nn\\
d(\tau, z) &= \tau^{-1/3} \left( d_0(v) + d_1(v) \frac{1}{\tau^{2/3}} + ...\right)
\label{latetime}
\end{align}
and similarly for $b(\tau, z)$ and $c(\tau, z)$. Inserting the ansatz (\ref{ansatzmetric}) and 
(\ref{ansatzA}) with (\ref{latetime}) into
the combined system of Einstein-Maxwell and covariant Maxwell equations (\ref{EM}) and  (\ref{EM2})
will turn the equation of motions into a system of nonlinear ordinary differential equations
for the coefficients $a_i, ..., d_i$ $(i \geq 0)$. In principle, this system can then be solved order by
order in $\tau^{-2/3}$.

\subsection{Zeroth-order solution}

In the following we restrict to give an exact solution for the zeroth-order coefficients
$a_0(v), ..., d_0(v)$. The non-vanishing components of the Einstein-Maxwell equations are
\begin{equation}
\label{einsteinzero}
\begin{array}{l l}
(\tau\tau): & 4e^{-{a_0}(v)} v^3 {d_0}'(v)^2=6 {b_0}'(v)-v {b_0}'(v)^2+12 {c_0}'(v)\\
&~~~~~~~~~~~~~~~~~~~~~~~~~-2 v {b_0}'(v) {c_0}'(v)-3 v {c_0}'(v)^2-2 v {b_0}''(v)-4 v {c_0}''(v) \,,\\&\\
(y y): & 4e^{-{a_0}(v)} v^3 {d_0}'(v)^2=-6 {a_0}'(v)+v {a_0}'(v)^2-12 {c_0}'(v)\\
&~~~~~~~~~~~~~~~~~~~~~~~~~+2 v {a_0}'(v) {c_0}'(v)+3 v {c_0}'(v)^2+2 v{a_0}''(v)+4 v {c_0}''(v)\,,\\&\\
(\perp\perp): & 4e^{-{a_0}(v)} v^3 {d_0}'(v)^2=-6 {a_0}'(v)+v {a_0}'(v)^2-6 {b_0}'(v)+v {a_0}'(v) {b_0}'(v)+v {b_0}'(v)^2\\
&~~~~~~~~~~~~~~~~~~~~~~~~~-6 {c_0}'(v)+v {a_0}'(v) {c_0}'(v)+v{b_0}'(v) {c_0}'(v)+v {c_0}'(v)^2\\
&~~~~~~~~~~~~~~~~~~~~~~~~~+2 v {a_0}''(v)+2 v {b_0}''(v)+2 v {c_0}''(v)\,,\\&\\
(zz): & 4e^{-{a_0}(v)} v^3 {d_0}'(v)^2=6 a_0'(v)+6 b_0'(v)-v a_0'(v) b_0'(v)\\
&~~~~~~~~~~~~~~~~~~~~~~~~~+12 c_0'(v)-2 v a_0'(v) c_0'(v)-2 v b_0'(v)c_0'(v)-v c_0'(v)^2\,, \\&\\
(z\tau): & 6 a_0'(v)-4 b_0'(v)-v a_0'(v) b_0'(v)+v b_0'(v)^2+4 c_0'(v)\\
&~~~~~~~~~~~~~~~~~~~~~~~~~-2 v a_0'(v) c_0'(v)+2 v c_0'(v)^2+2v b_0''(v)+4 v c_0''(v)=0 \,.
\end{array}
\end{equation}
At zeroth order, the $z$- and $\tau$-components of the Maxwell equation both lead to the same
equation,
\begin{align}
\label{maxwellzero}
\left(2+v {a_0}'(v)-v {b_0}'(v)-2 v {c_0}'(v)\right) {d_0}'(v)=2 v {d_0}''(v)\,.
\end{align}
The other components are zero.

These equations can be simplified a lot. Note that only four out of the five plus one
equations are independent. We also find from a linear combination of the $\tau\tau$- , $zz$- and
$z\tau$-components of the Einstein-Maxwell equations that $b_0(v) = c_0(v)$. Next, the Maxwell
equation (\ref{maxwellzero}) can be solved for $d_0(v)$,
\begin{align}
\label{d0v}
d_0(v)=S_4 \int\limits_0^v \tilde v e^{\frac{1}{2}\left( a_0(\tilde v) - b_0(\tilde v) - 2c_0(\tilde v) \right)} \,d\tilde v,
\end{align}
where $S_4$ is some integration constant which will be fixed below.

Substituting this back into the Einstein equations, the two remaining independent
equations are given by the $\tau\tau$- and $zz$-components. The first one ($\tau\tau$) is an
equation for $b_0(v)$,
\begin{align}
\label{einsteinb_0}
\ddd 3 v b_0''(v) + 3 v (b_0'(v))^2 - 9 b_0'(v) + 8 q_0^2 v^5 e^{-3b_0(v)} = 0\,.
\end{align}
while the second one ($zz$),
\begin{align}
\label{einsteina_0}
\ddd a_0'(v) = -v \frac{(b_0'(v))^2 + 2 b_0''(v)}{2 - v b_0'(v)}\,,
\end{align}
can be used to find $a_0(v)$ as soon as a solution for $b_0(v)$ is known. Our primary goal
will be to solve (\ref{einsteinb_0}) for $b_0(v)$. $a_0(v)$ and $d_0(v)$ can then easily be
 obtained from (\ref{einsteina_0}) and (\ref{d0v}).

Later, in order to fix some integration constants, we will need the asymptotic solution
close to the boundary which can be expanded in powers of $v$ as (here we present it up to
${\cal O}(v^{10})$)
\begin{equation}
\label{zero_order_FG}
\begin{array}{l}
\ddd a_0(v) = -\varepsilon_0 v^4+\frac{10 {q_0}^2}{9}v^6-\frac{\varepsilon_0^2}{18}v^8-\frac{2 {q_0}^2 \varepsilon_0}{45}v^{10} 
+ \cdots \,,\\\\
\ddd b_0(v) = c_0(v) = \frac{\varepsilon_0}{3}v^4-\frac{2 {q_0}^2}{9}v^6-\frac{\varepsilon_0^2}{18}v^8+\frac{14 {q_0}^2 \varepsilon_0}{135}v^{10} 
+ \cdots \,,\\\\
\ddd d_0(v) = q_0 v^2-\frac{q_0 \varepsilon_0}{3}v^6+\frac{2 q_0^3}{9}v^8+\frac{q_0 \varepsilon_0^2}{9}v^{10}
+ \cdots \,.
\end{array}
\end{equation}
It shows us that the solution exists and is uniquely fixed by parameters $\varepsilon_0$ and $q_0$. 
Comparing the expression (\ref{d0v}) with the boundary behaviour (\ref{zero_order_FG}), we may immediately fix the
integration constant $S_4$ as
\begin{align}
S_4 = 2q_0 \,.
\end{align}
We now solve (\ref{einsteinb_0}) for $b_0(v)$. By setting 
\begin{align}
b_0(v)=\log(\beta(v)) \,,
\end{align}
we simplify this equation to the form
\begin{align}
v \beta'' - 3\beta' + \frac{8}{3}q_0^2 v^5 \beta^{-2} = 0\,,
\end{align}
which turns out to be the modified Emden-Fowler equation \cite{Zaitsev}. 
Its solution can be
written in the parametric form
\begin{align}
\label{betaparametric}
\ddd\beta(v) = p S_3^2\exp\left\{  S_2\int\limits_{p_+}^p\left(
\frac{\tilde p^2}{4}+S_1+\frac{q_0^2}{3}\frac{1}{\tilde p} \right)^{-1/2} d \tilde p \right\}
\end{align} 
and
\begin{align}
\label{vparametric}
\ddd v = S_3\exp\left\{  \frac{S_2}{2}\int\limits_{p_+}^p\left(
\frac{\tilde p^2}{4}+S_1+\frac{q_0^2}{3}\frac{1}{\tilde p} \right)^{-1/2} d \tilde p \right\}\,.
\end{align} 
Here $S_1$, $S_2$, $S_3$ and $p_+$ are some integration constants. One can in principle absorb $p_+$ in $S_3$ but we separate them for the moment.
There are two useful expressions for $\beta(v)$ and $\beta'(v)$,
\begin{align}
\label{betasimple}
\beta(v) = p v^2
\end{align}
and
\begin{align}
\label{betaderivative}
\ddd\frac{d\beta(v)}{d v} = \frac{2 v}{S_2}\sqrt{\frac{p^2}{4}+S_1+\frac{q_0^2}{3}\frac{1}{p}} 
+ 2 p v\,.
\end{align}

From (\ref{zero_order_FG}), we get the near-boundary conditions
\begin{align}
\label{betaboundary}
\beta(v) = 1 + O(v^4)\,, \qquad \beta'(v) = \frac{4\varepsilon_0}{3}v^3 + O(v^5)\,,
\end{align}
which will be used to fix the integration constants $S_1$ and $S_2$. 

Comparing (\ref{betasimple}) with (\ref{betaboundary}), we find that near the boundary $v$
should behave as
\begin{align}
\label{asympt1}
v \approx \frac{1}{\sqrt{p}} \,,
\end{align}
which is small if $p$ is large. This should be compared with the general large-$p$
behaviour
\begin{align}
\label{asympt2}
v = S_3 \exp\left\{  \frac{S_2}{2} \int\limits^{p}_{p_+}\left(\frac{\tilde p}{2} + \cdots \right)^{-1} d \tilde p  \right\}\approx {const.}\ p^{S_2} \,.
\end{align}
This fixes $S_2$ as
\begin{align}
\label{s2}
S_2 = -\frac{1}{2}\,.
\end{align}

Substituting (\ref{asympt1}) into (\ref{betaboundary}), we extract the expected asymptotics 
for $\beta'(v)$ as a function of $p$,
\begin{align}
\label{asympt3}
\beta'(v) \approx \frac{4\varepsilon_0}{3}\frac{1}{p^{3/2}} \,.
\end{align}
Generically, at large $p$, (\ref{betaderivative}) is approximated by
\begin{align}
\beta'(v) = -4 v \sqrt{S_1 + \frac{p^2}{4} + \cdots} + 2 p v \approx -\frac{4 S_1}{p^{3/2}}\,,
\end{align}
which fixes $S_1$ as
\begin{align}
\label{s1}
S_1 = -\frac{\varepsilon_0}{3}.
\end{align}
Remarkably both constants $S_1$ and $S_2$ do not depend on $q_0$.

Let us now relate $S_3$ and $p_+$ by setting $S_3=v_+$ with $v_+ \equiv v(p_+)$. 
$v_+$ will be fixed by the requirement that the outer horizon of the
geometry is located at $v=v_+$. Formally, the horizon $v_+$ is defined as the
largest zero of the denominator on the right hand side of (\ref{einsteina_0}), 
\begin{align}
2 - v_+ \frac{\beta'(v_+)}{\beta(v_+)} = 0\,.
\end{align} 
This can be rewritten in terms of $p_+$. Using (\ref{betaderivative}), (\ref{betasimple}) and the expressions for $S_1$ and $S_2$, we get the condition
\begin{align}
\label{pequation}
p_+^3 - \frac{4\varepsilon_0}{3}p_+ + \frac{4 q_0^2}{3} =0 \,,
\end{align}
which can be solved by Cardano's formula. The largest solution of this equation is\footnote{{
In order to extract the roots correctly we use the following standard convention: 
\newline $a_+ a_- = \left(\frac{2}{3}\right)^{4/3}\varepsilon_0$, where 
$a_\pm=\left(-q_0^2 \pm\sqrt{q_0^4-\frac{16}{81}\varepsilon_0^3} \right)^{1/3}$.  }}
\begin{align}\label{pplus}
p_+ = \left(\frac{2}{3}\right)^{1/3}\left(\left(-q_0^2+\sqrt{q_0^4-\frac{16}{81}\varepsilon_0^3} \right)^{1/3}+\left(-q_0^2-\sqrt{q_0^4-\frac{16}{81}\varepsilon_0^3} \right)^{1/3}\right) \,.
\end{align}

The last step is to fix $v_+$ in (\ref{vparametric}). This can be done by substituting
the $p_+$ solution (\ref{pplus}) (and all the constants $S_{1,2,3}$) back into 
(\ref{vparametric}) and expand $v(p)$ for large $p$. In this way we determine the
constant on the right hand side of (\ref{asympt2}) as a function of $v_+$. Since this
constant must be one, we get
\begin{align}
\label{vplus}
v_+(\varepsilon_0,q_0) = \exp\left\{{\ddd\frac{1}{2}\lim\limits_{p\rightarrow\infty}\left[\int\limits^p_{p_+}\frac{d\tilde p}{\sqrt{4U(\tilde p)}}-\log p\right]}\right\} \,,
\end{align}
where $U(\tilde p)$ is defined as
\begin{align}\label{U}
U(\tilde p)= \frac{\tilde p^2}{4}-\frac{\varepsilon_0}{3} + \frac{q_0^2}{3\tilde p} \,.
\end{align}
For $q_0 = 0$, the integral can be performed analytically and $v_+$ 
reduces to the well-known result for the horizon \cite{Janik2005}, 
\begin{align}
v_+ (\varepsilon_0, q_0)\vert_{_{q_{_0}\rightarrow 0}} = \sqrt[4]{\frac{3}{\varepsilon_0}} \,.
\end{align}

\begin{figure}[!h]
	\centering
    {\includegraphics[width=8cm]{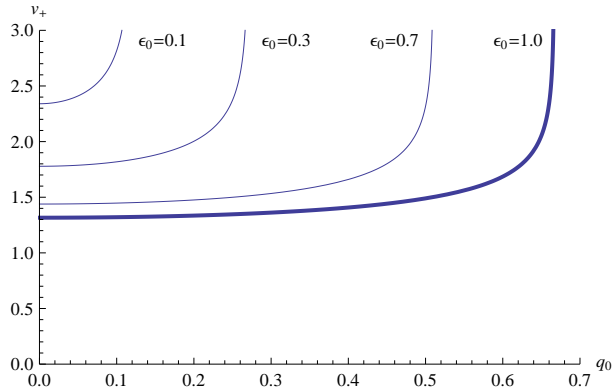}} 
\caption{\label{outer_horizon} The outer horizon $v_+ = v_+ (\varepsilon_0, q_0)$.}
\end{figure} 
For general $q_0$, this integral can in principle be written as a lengthy expression of
elliptic integrals of the first and third kind, $F(\phi, k)$ and $\Pi(n; \phi | m)$, respectively, which we will not do here. Instead, in Fig.~\ref{outer_horizon} we show the dependence of $v_+$ on the charge $q_0$ for some particular choices of $\varepsilon_0$. We note that for each 
$\varepsilon_0$ there is some maximal allowed value of the charge at which the black hole becomes extremal. This value can be found from the condition that the discriminant $\Delta$ of
(\ref{pequation}) vanishes,
\begin{align}
 \Delta = \left( -\frac{4}{9}\varepsilon_0 \right)^3 + \left( -\frac{2}{3}q_0^2 \right)^2 = 0  \,,
\end{align}
which leads to the bound
\begin{align}
\label{boundFG}
q_0 \leq q_0^{extr.} = \frac{2}{3}\varepsilon_0^{3/4} \,.
\end{align}

\begin{figure}[t]
     \centering
     \subfigure[\label{a0_plot}]
     {\includegraphics[width=5cm]{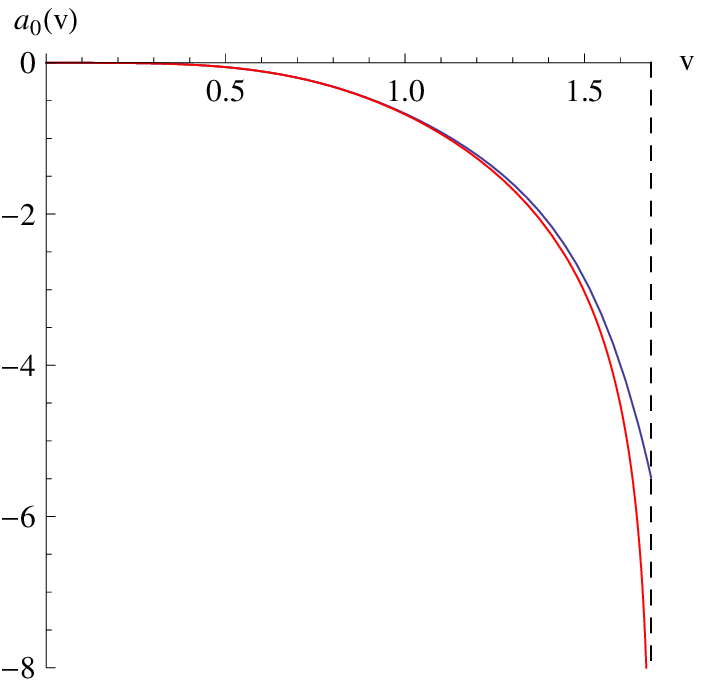}}\hspace{0cm}
     \centering
     \subfigure[\label{b0_plot}]
     {\includegraphics[width=5cm]{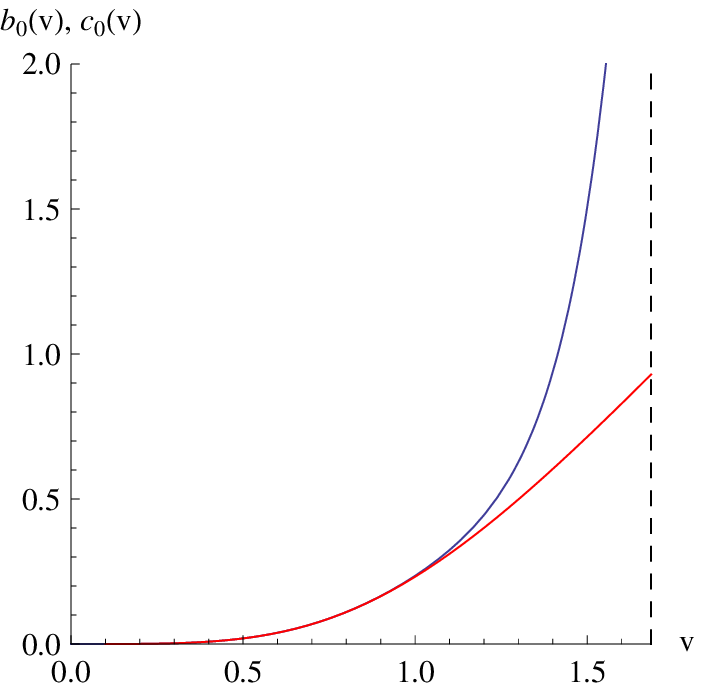}}\hspace{0cm}
     \centering
     \subfigure[\label{d0_plot}]
     {\includegraphics[width=5cm]{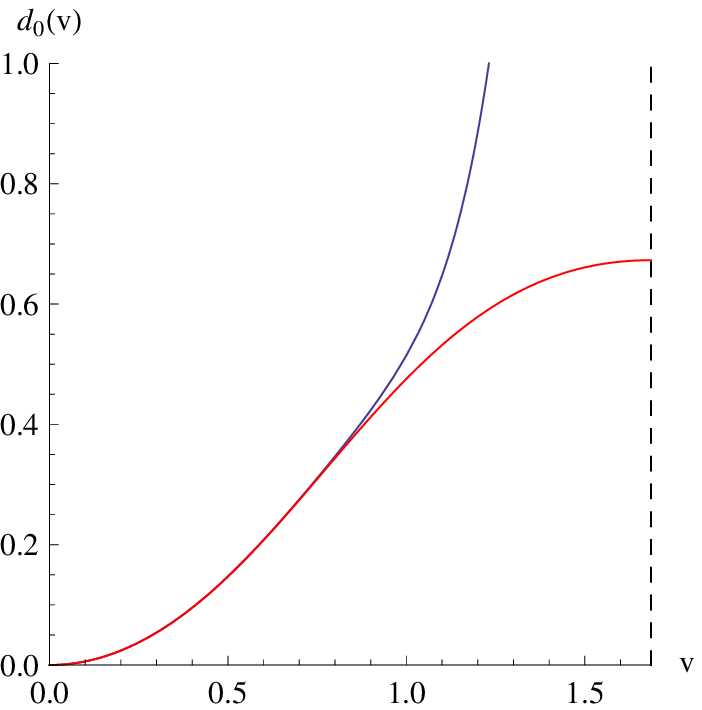}}\hspace{0cm}
\caption{\label{comparingplots} Exact solutions (red curves) and their near-boundary power expansions (blue curves) for $\varepsilon_0 = 1$ and $q_0 = 0.6\lesssim q_0^{extr.}$. The black dashed lines correspond to the horizon $v_+(\varepsilon_0, q_0) \simeq 1.685$.}
\end{figure}

In Fig.~\ref{comparingplots} we present some plots of the exact solution and compare them with the power expansions (\ref{zero_order_FG}). For the particular choice $\varepsilon_0 = 1$ and $q_0 = 0.6\lesssim q_0^{extr.}$ the difference between both curves is clearly visible. The function
$a_0(v)$ by definition has a singularity on the horizon, as can be seen in 
Fig.\ref{a0_plot}. The other functions $b_0(v)$, $c_0(v)$ and $d_0(v)$ are regular on the horizon and their power expansions are valid up to $v \lesssim 1$. Note also that $d_0(v)$ grows quadratically near the boundary, which reflects the Coulomb law in $D = 5$ dimensions. Near the horizon it approaches some finite constant value $\mu_0$ related to the chemical potential as
\begin{align}
\mu = A_0|_{boundary}-A_0|_{horizon} = \frac{d_0(v_+) - d_0(0)}{\tau^{1/3}} = \frac{ \mu_0}{\tau^{1/3}} \,,
\end{align}
which confirms the scaling behaviour (\ref{T_and_mu_scaling}).

\medskip
In summary, the zeroth-order solution $b_0(v)$ is given by
\begin{align}
e^{b_0(v)}&= p v_+^2 \exp \big(-\frac{1}{2} \int_{p_+}^p U(\tilde p)^{-1/2} d\tilde p\big)\,, \nn\\
v &=  v_+  \exp\big(-\frac{1}{4} \int_{p_+}^p U(\tilde p)^{-1/2} d\tilde p\big) \label{bsoln}\,,
\end{align}
with $v_+$ and $U(\tilde p)$ given by (\ref{vplus}) and
(\ref{U}), respectively. $a_0(v)$ and $d_0(v)$ are obtained by substituting $b_0(v)$
in (\ref{einsteina_0}) and (\ref{d0v}).

\subsection{Fefferman-Graham vs.\ Eddington-Finkelstein coordinates}\label{sec34}
The zeroth-order solution in Fefferman-Graham (FG) coordinates  can be related to that
in Eddington-Finkelstein (EF) coordinates by the coordinate transformation
\begin{align}
\label{EFtoFG}
\tilde \tau=\tau  \,,\qquad r=\frac{1}{z} e^{b(\tau, z)/2}\,.
\end{align} 
Transforming the Eddington-Finkelstein metric (\ref{ansatz})-(\ref{endofansatz}) 
and comparing the result with the power expansion (\ref{zero_order_FG}), we find  
\begin{align}
\label{q_relation}
q_0 = \frac{\sqrt{3}}{2} \tilde q_0.
\end{align}

Comparing also the bound (\ref{boundFG}) with that in Eddington-Finkelstein coordinates given by (\ref{boundEF}),
we find some relation between $\varepsilon_0$ and $m_0$ using (\ref{q_relation}):
\begin{align}\label{id}
 \varepsilon_0 = \frac{3}{4} m_0 \,.
\end{align}
For $q_0=0$ this relation can be easily checked by our solution with the general form of the metric 
in \cite{Janik2005, Nakamura, Hellerthesis}. 
In this case, we find that our solution (\ref{bsoln}) reduces to 
\begin{align}
e^{b_0(v)} = 1 + \frac{\varepsilon_0}{3} v^4  \,,
\end{align}
and similarly $a_0(v)$, such that
\begin{align}
\label{hellermetric}
\left.ds^2\right\vert_{q_0=0}= \frac{1}{z^2}\left[dz^2-\frac{(1-\frac{m_0}{4} v^4)^2}{1+\frac{m_0}{4} v^4} d\tau^2
+\tau^2 \left(1+\frac{m_0}{4} v^4\right) dy^2 + \left(1+\frac{m_0}{4} v^4\right) dx_\perp^2\right] \,.
\end{align}

\begin{figure}[!h]
	\centering
    {\includegraphics[width=7cm]{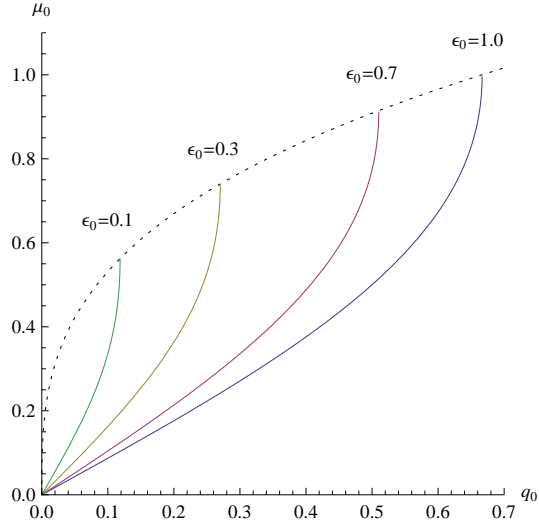}} 
\caption{\label{muplot} The chemical potential $ \mu_0 =  \mu_0(\varepsilon_0, q_0)$. The dotted line corresponds to the upper bound for $\mu_0$.}
\end{figure}

We also note that the transformation (\ref{EFtoFG}) relates the outer horizons, 
$\zeta_+$ in EF coordinates and $p_+$ in FG coordinates, as $\zeta_+ = \sqrt{p_+}$.
The chemical potential $\mu_0$ can therefore be written as a function of $\varepsilon_0$ 
and $q_0$. Using (\ref{m_definition}) and (\ref{q_relation}), we find
\begin{align}
 {\mu_0}(\varepsilon_0, q_0) = \tmu_0 = \frac{\sqrt{3}}{2}\frac{\tilde q_0}{\zeta_+^2} 
 =   \ddd\frac{q_0}{p_+(\varepsilon_0, q_0)}\,.
\end{align}
This dependence is shown in Fig.~\ref{muplot} for some particular values of $\varepsilon_0$.

Using this expression with the definition (\ref{pequation}) for $p_+$ and substituting there the maximal value for $q_0$ (\ref{boundFG}) we  find the following bound for the chemical potential:
\begin{align}
\label{mubound}
\mu_0(\varepsilon_0, q_0) \leq \mu_0^{extr.} = \left(\frac{3}{2}q_0^{extr.}\right)^{1/3} = \varepsilon_0^{1/4}\,.
\end{align}
This is not in contradiction with our earlier statement that the disappearance of the 
horzion does not impose a bound on
$\tilde \mu_0/\Lambda$. Note that if we identify 
$\varepsilon_0$ with $m_0$ as in (\ref{id}), then $\varepsilon_0$ explicitly
depends on $\tilde \mu_0=\mu_0$ and therefore (\ref{mubound}) is not a bound on
$\tilde \mu_0/\Lambda$.

\setcounter{equation}{0}
\section{Conclusions}

We constructed a natural extension of the late-time boost-invariant background
found in~\cite{Janik2005} (and \cite{Nakamura}--\cite{Hellerthesis}) to a background 
dual to an expanding $\N=4$ plasma {\em with} chemical potential.
The solution we found depends on two parameters, the 
chemical potential $\tmu_0$ and temperature scale $\Lambda$, which are encoded
in the mass parameter $m_0$ and charge $\tilde q_0$ of a time-dependent AdS 
Reissner-Nordstr\"om-like solution.
In Eddington-Finkelstein coordinates the first-order solution is given by the expansion (\ref{ansatz})--(\ref{endofansatz}),
with the zeroth-order and first-order coefficients given by (\ref{zerocoeff}) and
(\ref{firstcoeff}), respectively. We showed that the viscosity of the boundary theory
computed from the time-dependent solution
is in agreement with that in \cite{Erdmenger}.
 We also constructed a zeroth-order solution
in Fefferman-Graham coordinates, which we presented in parametric form,
see the general ansatz (\ref{ansatzmetric}) and (\ref{ansatzA}) with (\ref{bsoln}).
FG coordinates may be the preferred choice, when strings \cite{Kim:2007ut} or branes 
\cite{Grosse, Evans} are embedded into  the geometry.
Finally, we found the coordinate transformation which maps the zeroth-order
solution in FG coordinates to that in EF coordinates.

We argued in several ways that the charge density behaves like $\tau^{-1}$ at
all times.
Unlike the energy density, it can not be chosen freely at early times. This
is basically because the charge density behaves like $\tau^{-1}$ at large
$\tau$, see  {\em e.g.}~(\ref{chdensityasymp}) or (\ref{asymp}), and higher-order
corrections are absent. It also follows directly from the equations of motion,
see (\ref{chdensity}) which holds for all times. Naive extrapolation to early times shows a 
singularity in the gauge field at $\tau=0$. However, this does not signal a breakdown of the
hydrodynamic approximation since the total charge is constant at all times
and therefore regular even at $\tau=0$ (see footnote~\ref{footnote5} on p.~\pageref{footnote5}).

A possible application of the background, when appropriately modified and
extended, could be the 
chiral magnetic effect (CME) \cite{Warringa}. The CME states that, 
in the presence of a magnetic field and non-zero chiral chemical potential,
an electromagnetic current of the type 
${\mathbf J}\propto \mu_5 {\mathbf B}$ is generated in the plasma.
The CME is a non-equilibrium
process and requires the introduction of gauge fields with (time-dependent) spatial
components. For instance, for an electric field $E$, one needs to introduce
the spatial component $A_3=A_3(\tau, r)$ with boundary condition
$A_3 \rightarrow \tau E$ at $r \rightarrow \infty$. The time-dependence
of the gauge field reflects non-equilibrium physics and 
requires the back-reaction on the geometry, unless one keeps it infinitesimal \cite{Rebhan2009}
(see also \cite{Landsteiner, Rubakov} for an AdS/CFT approach to the CME). 
In this case one would also obtain higher-order corrections to the charge 
density due the effects of the Chern-Simons term, which are absent 
in our solution.
An attempt to include an $E$ field in the dual of
an electrified plasma (without chemical potential) has been made in 
\cite{Yee}.

Finally, it would be interesting to find a numerical solution of our background 
{\em \`a la} Chesler and Yaffe \cite{Chesler} which would hold beyond the hydrodynamic regime.

\subsection*{Acknowledgments}

We would like to thank Johanna Erdmenger, Nick Evans, Michael Haack, Romuald Janik,
Keun-young Kim and Volker Schomerus for helpful discussions related to this work.

\newpage
\appendix

\noindent {\LARGE \bf Appendix}

\setcounter{equation}{0}
\section{Roots of (\ref{roots})}\label{appA}

For completeness, we present the six roots of (\ref{roots}) in this appendix.
The equation (\ref{roots}) is depressed bicubic in $\zeta_i$ and, therefore, can be solved by
Cardano's formula. It has six solutions, which can be expressed as
\begin{align}
 \zeta_{i} \in \left\{ \pm\sqrt{\alpha_+ + \alpha_-}, \pm\sqrt{-\frac{\alpha_+ + \alpha_-}{2} - i\frac{\alpha_+ - \alpha_-}{2}\sqrt{3}}, \pm\sqrt{-\frac{\alpha_+ + \alpha_-}{2} + i\frac{\alpha_+ - \alpha_-}{2}\sqrt{3}} \right\}\,,
\end{align}
where
\begin{align}
\alpha_{\pm}^3 = -\frac{\tilde q_0^2}{2} \pm \sqrt{\frac{\tilde q_0^4}{4}-\frac{m_0^3}{27}} \,.
\end{align}
Here we use the standard convention $\alpha_{+}\alpha_{-} = m_0/3$.
One can recognize the outer horizon $\zeta_+\equiv r_+\tau^{1/3}$ in the first pair of solutions and the inner horizon $\zeta_-\equiv r_-\tau^{1/3}$ in the second one. 

\setcounter{equation}{0}
\section{The energy-momentum tensor (\ref{T_definition}) }
\label{extrinsic_curvature}

In this appendix we introduce the geometric quantities used for the computation of
the energy-momentum tensor (\ref{T_definition}).
Here we consider an $r=const.$\ four-dimensional surface with induced metric
$\gamma_{\mu\nu}$ on it:\footnote{Here and after all Greek letters denote a 5-index.}
\begin{eqnarray}
\label{gamma}
\gamma_{\mu\nu}=g_{\mu\nu}-{n}_{\mu}{n}_{\nu}\,,
\end{eqnarray}
where $g_{\mu\nu}$ is the 5-metric and ${n}^{\mu}$ is the outward-pointing unit normal vector to the surface. For our ansatz (\ref{ansatz}) it is given by 

\begin{align}
{n}_{\mu}&=\left(0,0,0,0,\frac{1}{\sqrt{-g_{\tilde\tau\tilde\tau}}}\right)\,,
\end{align}
where $g_{\tilde\tau\tilde\tau} = -r^2 e^{a(\tilde\tau,r)}$. The indices of the induced metric can be raised and lowered by means of the 5-metric $g_{\mu\nu}$,
\begin{align}
\gamma_\mu{}^\nu=\gamma_{\mu\alpha} g^{\alpha\nu}\,.
\end{align}

The surface extrinsic curvature is given by
\begin{align}
K_{\mu\nu} \equiv -\frac{1}{2}({}^{(4)}\nabla_{\alpha}{n}_{\beta}+{}^{(4)}\nabla_{\beta}{n}_{\alpha}) = -\frac{1}{2}\gamma_{\mu}^{\ \alpha}\gamma_{\nu}^{\ \beta} \left(\nabla_{\alpha}{n}_{\beta}+\nabla_{\beta}{n}_{\alpha}\right)\,,
\end{align}
where we put $^{(4)}$ to covariant derivatives associated with the induced metric,
while the derivatives on the right-hand side are defined with respect to the 5-metric.
We also define a scalar $K=K_{\mu\nu}g^{\mu\nu}=K_{\mu\nu}\gamma^{\mu\nu}$, which is used in the Gibbons-Hawking-York part of (\ref{T_definition}).

The Einstein tensor on the surface is defined as
\begin{eqnarray}
G_{\mu\nu}={}^{(4)}R_{\mu\nu}-\frac{1}{2}\gamma_{\mu\nu}{}^{(4)}R\,,
\end{eqnarray}
where the 4-tensors can be expressed through the 5-tensors (defined with respect to $g_{\mu\nu}$) by the Gauss equations:
\begin{eqnarray}
{}^{(4)}R{}^{\alpha}{}_{\mu\beta\nu} &=&
   R{}^{\kappa}{}_{\lambda\rho\sigma}
   \gamma^{\alpha}_{\ \kappa} \gamma_\mu{}^\lambda
   \gamma_\beta{}^\rho \gamma_\nu{}^\sigma 
   + K{}^{\alpha}_{\ \beta}K_{\mu\nu}-K_{\mu\beta}K{}^{\alpha}_{\ \nu},\\
{}^{(4)}R_{\mu\nu} &=&{}^{(4)}R^{\alpha}{}_{\mu\beta\nu}\gamma{}_{\alpha}^{\ \beta}=
\gamma_{\kappa}{}^{\lambda}\gamma^{\ \rho}_{\mu}\gamma^{\ \sigma}_{\nu}\:R^{\kappa}{}_{\rho \lambda\sigma}
+K K_{\mu\nu}-K_{\mu\alpha}K^{\alpha}_{\ \nu},  \\
{}^{(4)}R&=&{}^{(4)}R_{\mu\nu}\gamma^{\mu\nu}
=R-{n}^{\alpha}{n}^{\beta}\:R_{\alpha\beta}
+K^{2}-K_{\alpha\beta}K^{\alpha\beta},
\end{eqnarray}
where the raising/lowering rule is given by $K_{\nu}{}^\mu = K_{\nu\alpha}\gamma^{\alpha\mu} = \gamma_{\nu\alpha} K^{\alpha\mu}$.


\end{document}